\documentclass[12pt]{iopart}

\usepackage{graphicx}
\begin{document}

\title[Global polarization of QGP in non-central $AA$ collisions]
{Global polarization of QGP in non-central heavy ion collisions at high energies}

\author{Zuo-tang Liang}

\address{Department of Physics, Shandong University, Jinan, Shandong 250100, China}
\ead{liang@sdu.edu.cn}
\begin{abstract}

Due to the presence of a large orbital angular momentum of the
parton system produced at the early stage of non-central heavy-ion
collisions, quarks and anti-quarks are shown 
to be polarized in the direction opposite to
the reaction plane which is determined by the impact-parameter and the beam momentum.
The global quark polarization via elastic scattering
was first calculated in an effective static potential
model, then using QCD at finite temperature with 
the hard-thermal-loop re-summed gluon propagator. 
The measurable consequences are discussed.
Global hyperon polarization from the hadronization of polarized quarks are predicted
independent of the hadronization scenarios.
It has also been shown that the global polarization of quarks and anti-quarks
leads also to spin alignment of vector mesons.
Dedicated measurements at RHIC are underway and some of the
preliminary results are obtained.
In this presentation, 
the basic idea and main results of global
quark polarization are presented.
The direct consequences such as global hyperon polarization
and spin alignment are summarized.

\end{abstract}


\section{Introduction}

We all know that spin is a basic degree of freedom of the elementary particles.
Spin effects in high energy reactions usually provide us
with useful information on the reaction mechanism and often give us surprises.
Such effects have been studied intensively in high energy
lepton-hadron, hadron-hadron, and hadron-nucleus collisions
and lead to an active field of High Energy Spin Physics.
In contrast, less study has been made in this direction in high energy heavy-ion collisions.
One of the reasons might be that it would be very difficult 
or even impossible to polarize a heavy ion beam.
Recent studies [1-3] have shown that hadrons can be polarized 
w.r.t. the reaction plane in high energy 
$AA$ collisions with unpolarized beams. 
This is one of the places where we can study spin effects without 
a polarized beam and I hope that my talk can serve as an example 
which shows you, by looking at the spin degree of freedom, 
one can obtain some interesting information on the reaction mechanisms, 
even in heavy ion collision.

\section{Global orbital angular momentum and shear flow}

\begin{figure}[htbp]
\begin{center}
\includegraphics[width=5cm]{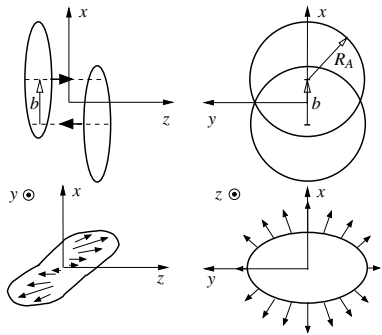}
\caption{Illustration of non-central $AA$ collision with
impact parameter $\vec{b}$.
The global angular momentum of the produced matter is along $-\hat{y}$,
opposite to the reaction plane.}
\label{fig1}
\end{center}
\end{figure}

We consider two colliding nuclei with the projectile of
beam momentum $\vec p_{in}$ moving in the direction of the $z$ axis, as
illustrated in Fig.~\ref{fig1}.
The impact parameter $\vec{b}$ 
is taken as $\hat{x}$-direction.
The normal $\vec{n}_b\propto \vec p_{in} \times \vec{b}$ 
of the reaction plane is taken as $\hat{y}$.
For a non-central $AA$ collision,
the dense matter system in the overlapped region 
will carry a global orbital angular momentum $L_y$
in the direction $-\hat{y}$.
The magnitude of $L_y$ is estimated 
using a hard spherical distribution for nucleus and is given in Fig. 2a. 
We see that $-L_y$ is indeed huge and is
of the order of $10^5$ at most $b$'s.

\begin{figure}[htbp]
\begin{center}
\includegraphics[width=3.8cm]{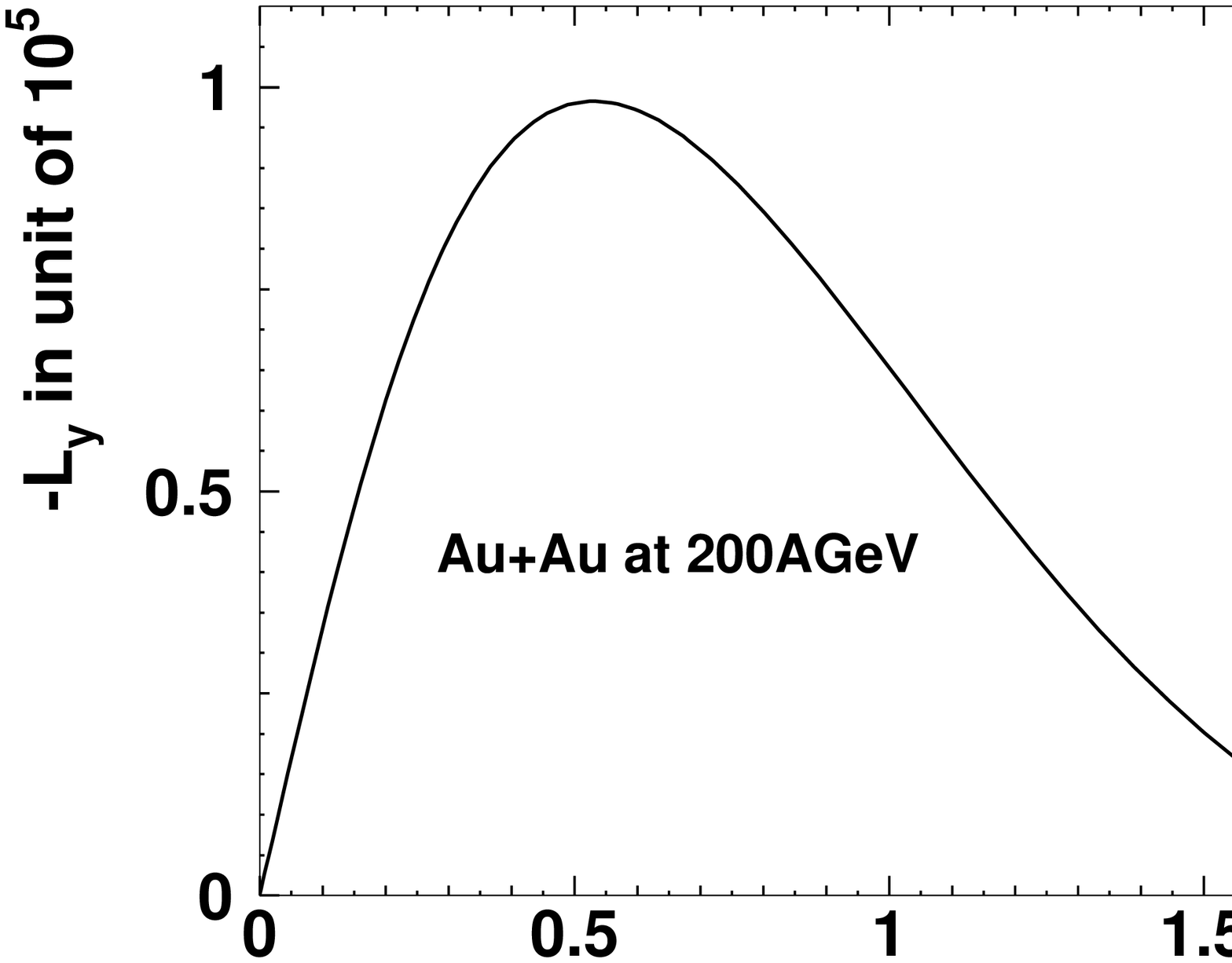}
\includegraphics[width=3.8cm]{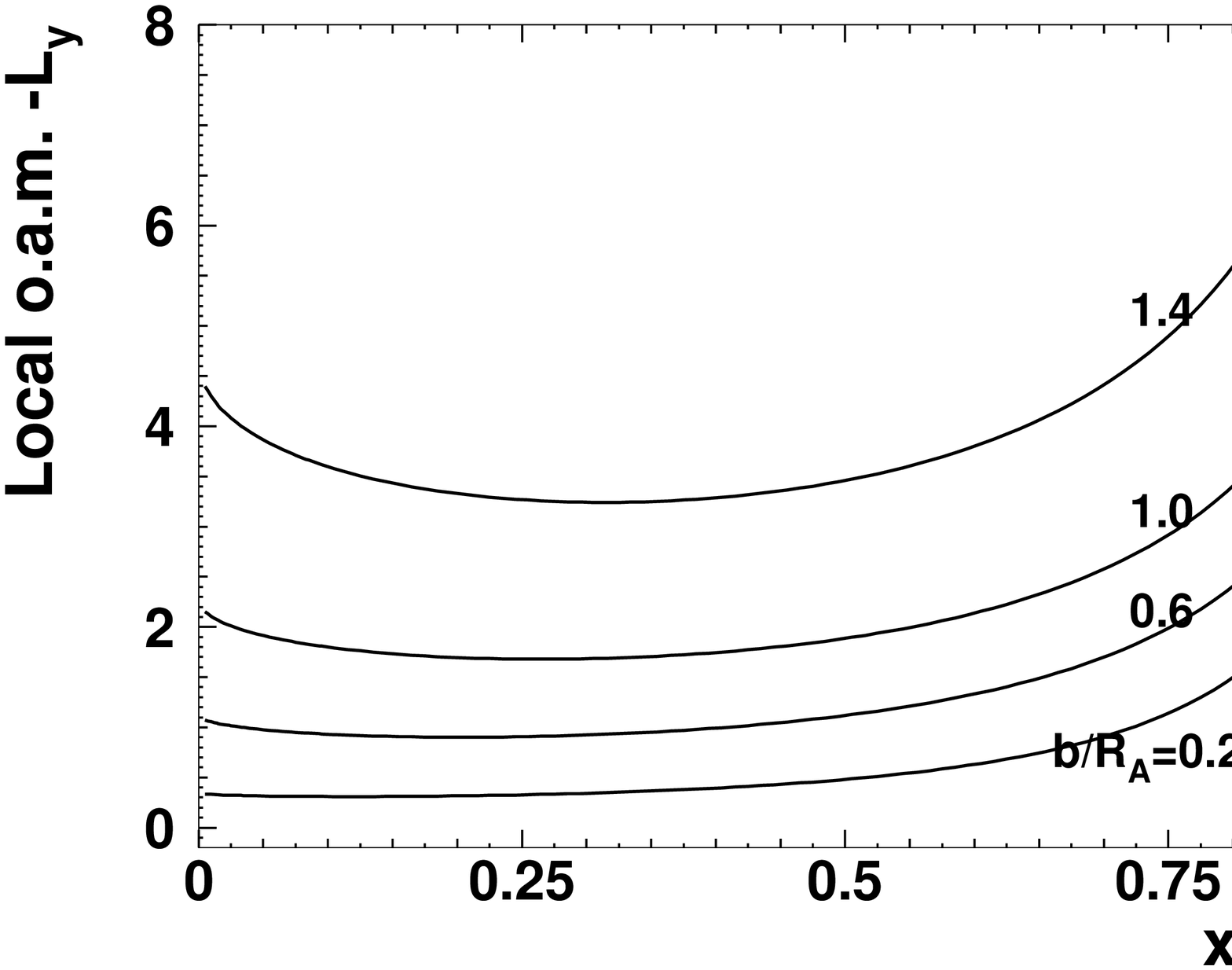}
\caption{a (left), The global orbital angular momentum $L_y$ of the overlapping system in a non-central
$AA$ collision at RHIC energy as a function of the impact parameter $b$; 
b (right), The average orbital angular momentum of two neighboring partons
separated by $\Delta x=1$fm as a function of $x/(R_A-b/2)$ for
different values of $b/R_A$.}
\end{center}
\end{figure}

Assuming that a partonic system is formed immediately after
the initial collision, interactions among the produced partons will
lead to the formation of a quark-gluon plasma (QGP) with both
transverse (in $x$-$y$ plane) and longitudinal collective motion.
The existence of the global orbital angular momentum of the system
discussed above implies a finite
transverse (along $\hat{x}$) gradient of the longitudinal flow velocity.

The initial collective longitudinal momentum can be calculated
as the total momentum difference between participant projectile
and target nucleons.
Since the measured total multiplicity in $AA$ collisions
is proportional to the number of participant nucleons\cite{phobos2},
we can assume the same for the produced partons
with a proportionality constant $c(s)$ at a given
center of mass energy $\sqrt{s}$.
Hence, the average collective
longitudinal momentum per parton is given by,
\begin{equation}
p_z(x,b;\sqrt{s})=\frac{\sqrt{s}}{2c(s)}
\frac{dN_{\rm part}^P/dx - dN_{\rm part}^T/dx}
{dN_{\rm part}^P/dx + dN_{\rm part}^T/dx}.
\end{equation}
$p_z(x,b;\sqrt{s})$ is a monotonically increasing function of $x$.
This can be seen more clearly by looking at the derivative $dp_z/dx$, 
which is almost a constant for different $x$. 
From $dp_z/dx$, we can estimate the average longitudinal
momentum difference $\Delta p_z$ between two neighboring partons 
separated by a transverse interval.
On the average, the relative orbital angular momentum
for two partons separated by $\Delta x$ in the transverse direction
is given by $l_y = -(\Delta x)^2dp_z/dx$.
E.g., for $Au+Au$ at $\sqrt{s}=200$ GeV, $c(s)\simeq 45$, $l_y$ for $\Delta x=1$fm is shown in Fig.~2b. 
We see that $l_y$ is in general of the order of 1 and is larger than the spin of a quark.
This implies that the effect can indeed be very significant.

We emphasize that the results given in Fig. 2 correspond to the average over rapidity, and in principle, 
we should consider the distribution of such collective longitudinal momentum over rapidity. 
This kind of distribution has been discussed e.g. by Adil and Gyulassy in 
their study on jet tomography of twisted QGP[5], and can also be calculated e.g. using HIJING. 
If we take this distribution into account and consdier that partons with slightly different rapidities 
can also interact with each other, we will still obtain a non-vanishing vorticity for the interacting parton system. 
Here, the final relevant quantity should be the local (both in transverse separation $x$ and in rapidity $\eta$) 
derivative of the longitudinal momentum distribution. 
The average results presented above can only be serve as a guide for the magnitude of this effect. 
In the following, we will discuss the polarization effects caused by such local vorticity.

\section{Global quark polarization w.r.t. the reaction plane}

To see whether the local orbital angular momentum between 
the neighboring partons in a QGP can be converted to quark polarization 
via parton scattering, we consider quark scattering at fixed impact parameter.
For definiteness, we consider a non-identical quark-quark scattering
$q_1(P_1,\lambda_1)+q_2(P_2,\lambda_2)\to q_1(P_3,\lambda_3)+q_2(P_4,\lambda_4)$,
where $P_i=(E_i,\vec p_i)$ and $\lambda_i$ denote
the 4-momentum and spin of the quark respectively.
We start with the usual cross section in momentum space,
\begin{equation}
\hspace{-2cm}
d\sigma_{\lambda_3}=\frac{c_{qq}}{F}\frac{1}{4}\sum_{\lambda_1,\lambda_2,\lambda_4}
\mathcal{M}(Q){\mathcal{M}}^{*}(Q)
(2\pi)^{4}\delta(P_1+P_2-P_3-P_4)\frac{d^3{\vec{p}}_3}{(2\pi)^{3}2E_3}
\frac{d^3{\vec{p}}_4}{(2\pi)^{3}2E_4},
\end{equation}
where ${\cal{M}}(Q)$ is the scattering amplitude in momentum space,
$Q=P_3-P_1=P_2-P_4$ is the 4-momentum transfer,
$c_{qq}=2/9$ and $F$ are the color and flux factors, respectively. 
The cross section in impact parameter space
is obtained by making a two dimensional Fourier transformation of
the transferred transverse momentum  $\vec q_T$, i.e.,
\begin{equation}
\frac{d\sigma_{\lambda_3}}{d^{2}{\vec{x}}_{T}}
=\frac{c_{qq}}{16F}\sum_{\lambda_1,\lambda_2,\lambda_4}
\int\frac{d^{2}{\vec{q}}_{T}}{(2\pi)^{2}}\frac{d^{2}{\vec{k}}_{T}}{(2\pi)^{2}}
e^{i({\vec{k}}_{T}-{\vec{q}}_{T})\cdot{\vec{x}}_{T}}
\frac{\mathcal{M}({\vec{q}}_{T})}{\Lambda({\vec{q}}_{T})}
\frac{{\mathcal{M}}^{*}({\vec{k}}_{T})}{{\Lambda}^{*}({\vec{k}}_{T})},
\end{equation}
where 
${\mathcal{M}}({\vec{q}}_{T})$ and ${\mathcal{M}}({\vec{k}}_{T})$
are the scattering matrix elements in momentum space
with 4-momentum transfer $Q=(0,\vec{q})$ and $K=(0,\vec{k})$ respectively,
$\Lambda({\vec{q}}_{T})=\sqrt{(E_1+E_2)|p+q_z|}$ is a kinematic factor.
The differential cross section can be divided into a
spin-independent and a spin dependent part, i.e.,
\begin{equation}
\frac{d\sigma_{\lambda_3}}{d^{2}{\vec{x}}_{T}}=
\frac{d\sigma}{d^{2}{\vec x}_{T}}
+\lambda_3\frac{d\Delta\sigma}{d^{2}{\vec x}_{T}}
\end{equation}
Parity conservation demands that  
they have the following form,
\begin{equation}
\frac{d\sigma}{d^{2}{\vec x}_{T}}=F(x_T,\sqrt{\hat{s}}), \phantom{XXX}
\frac{d\Delta\sigma}{d^{2}{\vec x}_{T}}
=\vec{n}\cdot({\vec{x}}_T\times{\vec{p}}\ )\Delta F(x_T,\sqrt{\hat{s}}),
\end{equation}
where $\vec{n}$ is the polarization vector for $q_1$ in its rest frame.
$F(x_T,\sqrt{\hat{s}})$ and $\Delta F(x_T,\sqrt{\hat{s}})$ are both functions of
 $x_T\equiv|\vec{x}_T|$ and the energy $\sqrt{\hat{s}}$ of the quark-quark system.
This is because, in an unpolarized reaction,
the cross section should be independent of any transverse direction.
For the spin-dependent part, the only scalar that we can construct from
the vectors that we have at hand is $\vec n\cdot(\vec p\times\vec x_T)$.

We note that,
$\vec{x}_T\times\vec{p}$ is nothing else but the
relative orbital angular momentum of the $q_1q_2$-system, i.e.,
$\vec{x}_T\times\vec{p}=\vec{l}$.
We see from Eq.(5) that the cross section takes its maximum
$\vec n$ is parallel to $\vec l$ or $-\vec l$
depending on whether $\Delta F$ is positive or negative.
This corresponds to a polarization of quark in the
direction $\vec l$ or $-\vec l$.

As discussed in last section, for $AA$ collisions with given reaction plane,
the direction of the averaged $\vec l$ of the two scattered quarks is given.
Since a given direction of $\vec l$ corresponds to a given direction
of $\vec x_T$, this implies that there should be a preferred direction
of $\vec x_T$ over others at a given direction of $\vec b$.
The detailed distribution of $\vec x_T$ at given $\vec b$
depends on the collective longitudinal momentum distribution discussed above.
For simplicity, we considered an uniform distribution of
$\vec x_T$ in all the possible directions
in the half $oxy$-plane with $x>0$.
In this case, we need to
integrate $d\sigma/d^{2}\vec x_T$ and $d\Delta\sigma/d^{2}\vec x_T$
in the half plane to obtain the average cross section at a given $\vec b$, i.e.,
\begin{equation}
\sigma=\int_{0}^{+\infty}dx\int_{-\infty}^{+\infty}dy\quad\frac{d\sigma}{d^{2}{\vec{x}}_{T}},
\phantom{XXX}
\Delta\sigma =
\int_{0}^{+\infty}dx\int_{-\infty}^{+\infty}dy\quad\frac{d\Delta\sigma}{d^{2}{\vec{x}}_{T}},
\label{dcsaverage}
\end{equation}
The polarization of the quark after one scattering is given by,
$P_q=\Delta\sigma/\sigma$.

\subsection{Results under small angle approximation}

The calculations are in principle straight forward but in practice
very much complicated.
Hence, in Ref.[1], we have given an example by
calculating them using a screened static potential model
and in the ``small angle approximation''.
The results are given by,
\begin{eqnarray}
\Bigl[\frac{d\sigma}{d^{2}{\vec x}_{T}}\Bigr]_{SPM}&=&\frac{g^4c_{T}}{2}
\int\frac{d^{2}{\vec{q}}_{T}}{(2\pi)^{2}}\frac{d^{2}{\vec{k}}_{T}}{(2\pi)^{2}}
\frac{e^{i({\vec{k}}_{T}-{\vec{q}}_{T})\cdot\vec{x}_T}}
{(q_T^{2}+\mu_D^2)(k_T^{2}+\mu_D^2)}\\
\Bigl[\frac{d\Delta{\sigma}}{d^{2}{\vec x}_{T}}\Bigr]_{SPM}
&=&-i\frac{g^4
c_{T}}{4} \int\frac{d^{2}{\vec q}_{T}}{(2\pi)^{2}}\frac{d^{2}{\vec k}_{T}}{(2\pi)^{2}}
\frac{(\vec{k}_T-\vec{q}_T)\cdot(\vec{p}\times\vec{n})e^{i(\vec{k}_T-\vec{q}_T)\cdot\vec{x}_T}}
{p^2(q_T^{2}+\mu_D^2)(k_T^{2}+\mu_D^2)}.
\end{eqnarray}
Carrying out the integrations over $\vec q_T$ and $\vec k_T$, we obtain that,
\begin{equation}
\Bigl[\frac{d\sigma}{d^{2}{\vec x}_{T}}\Bigr]_{SPM}
=2\alpha_s^2c_T 
\frac{1}{(2\pi)^2}K_0^2(\mu_{D}x_T),
\end{equation}
\begin{equation}
\Bigl[\frac{d\Delta\sigma}{d^{2}{\vec x}_{T}}\Bigr]_{SPM}
=\alpha_s^2c_T\mu_D 
\frac{(\vec{p}\times\vec{n})\cdot\hat{\vec{x}}_T}{p^2}
K_0(\mu_{D}x_T)K_1(\mu_{D}x_T)),
\end{equation}
where $J_0$ and $K_0$ are the Bessel and modified Bessel functions respectively. 
Carrying out the integrations in the half plane with $x>0$ and we obtained that[1],
\begin{equation}
P_q=-\pi\mu_D p/ 2E(E+m_q).
\end{equation}
The result is very encouraging since it shows a quite significant negative polarization 
of the quark after one scattering.

More accurate calculations should be made 
using QCD at finite temperature. 
The quark-quark scattering 
is described by a Hard-Thermal-Loop (HTL) re-summed
gluon propagator~\cite{screen},
\begin{equation}
\Delta^{\mu \nu}(Q)=\frac{P_T^{\mu\nu}}{-Q^2+\Pi_T}
+\frac{P_L^{\mu\nu}}{-Q^2+\Pi_L}+(\alpha-1)\frac{Q^{\mu}Q^{\nu}}{Q^4},
\end{equation}
where $Q$ is the gluon four momentum, 
$\alpha$ is a gauge fixing parameter, 
\begin{equation}
  {P}_L^{\mu\nu}=\frac{-1}{Q^2q^2}(\omega Q^{\mu}-Q^2U^{\mu})
  (\omega Q^{\nu}-Q^2U^{\nu})\, , \ \ \ 
  P_T^{\mu\nu}=\tilde{g}^{\mu\nu}
  +\frac{\tilde{Q}^{\mu}\tilde{Q}^{\nu}}{q^2}\, ,
\end{equation}
\begin{eqnarray}
\Pi_L&=&\mu_D^2\left[1-\frac{x}{2}
\ln\left(\frac{1+x}{1-x}\right) + i \frac{\pi}{2}x \right](1-x^2), \\
\Pi_T&=&\mu_D^2\left[\frac{x^2}{2}+\frac{x}{4}(1-x^2)
\ln\left(\frac{1+x}{1-x}\right) - i \frac{\pi}{4}x(1-x^2)\right],
\end{eqnarray}
where $\omega=Q\cdot U$, 
  $\tilde{Q}=Q-\omega U$, 
  $q^2=-\tilde{Q}^2$,
  $\tilde{g}_{\mu\nu}=g_{\mu\nu}-U_{\mu}U_{\nu}$, 
  $x=\omega/q$, $\mu_D^2=g^2(N_c+N_f/2)T^2/3$ is the Debye screening mass, 
  $U$ and $T$ are respectively the fluid velocity and temperature of heat bath.
In this framework, we have,
\begin{equation}
\mathcal{M}({\vec{q}}_{T})= {\overline{u}}_{{\lambda}_3}(P_1+Q){\gamma}_{\mu}
u_{{\lambda}_1}(P_1) {\Delta}^{\mu\nu}(Q)
{\overline{u}}_{{\lambda}_4}(P_2-Q){\gamma}_{\nu}u_{{\lambda}_2}(P_2),
\end{equation}
We work in the center of mass frame, and in Feynman gauge, so we have, 
\begin{equation}
{\Delta}^{\mu\nu}(Q)=
\frac{g^{\mu\nu}-U^\mu U^\nu}{q^2}
+\frac{U^\mu U^\nu}{q^2+{{\mu}_D}^2}
\end{equation}
The HTL gluon propagator needs to be regularized.
We do this by introducing a non-perturbative magnetic mass $\mu_m\approx0.255\sqrt{N_c/2} g^2T$~\cite{TBBM93} 
into the transverse self-energy. 
For simplicity, we consider only the longitudinal momentum distribution of the 
partons in QGP and have, in c.m. frame, $U^{\mu}=(1,0,0,0)$.

Under the ``small angle approximation'', we obtain that,
\begin{equation}
\hspace{-2cm}
\frac{d\sigma}{d^{2}{\vec x}_{T}}=\frac{g^4c_{qq}}{8}
\int\frac{d^{2}{\vec{q}}_{T}}{(2\pi)^{2}}\frac{d^{2}{\vec{k}}_{T}}{(2\pi)^{2}}
e^{i({\vec{k}}_{T}-{\vec{q}}_{T})\cdot\vec{x}_T}
(\frac{1}{q_T^2+\mu_m^2}+\frac{1}{q_T^2+\mu_D^2})
(\frac{1}{k_T^2+\mu_m^2}+\frac{1}{k_T^2+\mu_D^2}), 
\end{equation}
\begin{eqnarray}
&&\frac{d\Delta{\sigma}}{d^{2}{\vec x}_{T}}
=-i\frac{g^4c_{qq}}{16}
\int\frac{d^{2}{\vec q}_{T}}{(2\pi)^{2}}\frac{d^{2}{\vec k}_{T}}{(2\pi)^{2}}
e^{i(\vec{k}_T-\vec{q}_T)\cdot\vec{x}_T}
\frac{(\vec{k}_T-\vec{q}_T)\cdot(\vec{p}\times\vec{n})}{p^2}\nonumber\\
&&\phantom{XXXXXX}\times
(\frac{1}{q_T^2+\mu_m^2}+\frac{1}{q_T^2+\mu_D^2})
(\frac{1}{k_T^2+\mu_m^2}+\frac{1}{k_T^2+\mu_D^2}).
\end{eqnarray}
We see that the difference between the results obtained with HTL propagator and 
those in the static potential model is
the additional contributions from the magnetic part. 

\subsection{Numerical results}

The expressions for the cross sections without ``small angle approximation'' 
are quite complicated. I will not present them here. 
Interested readers are referred to [3]. 
We have carried out the integrations numerically and obtained 
the preliminary results as shown in Fig.~3a.  
For comparison, we show the results together with those obtained under 
the small angle approximation in Fig.~3b.

\begin{figure}[htbp]
\begin{center}
\hspace{1cm}
\includegraphics[width=6.5cm]{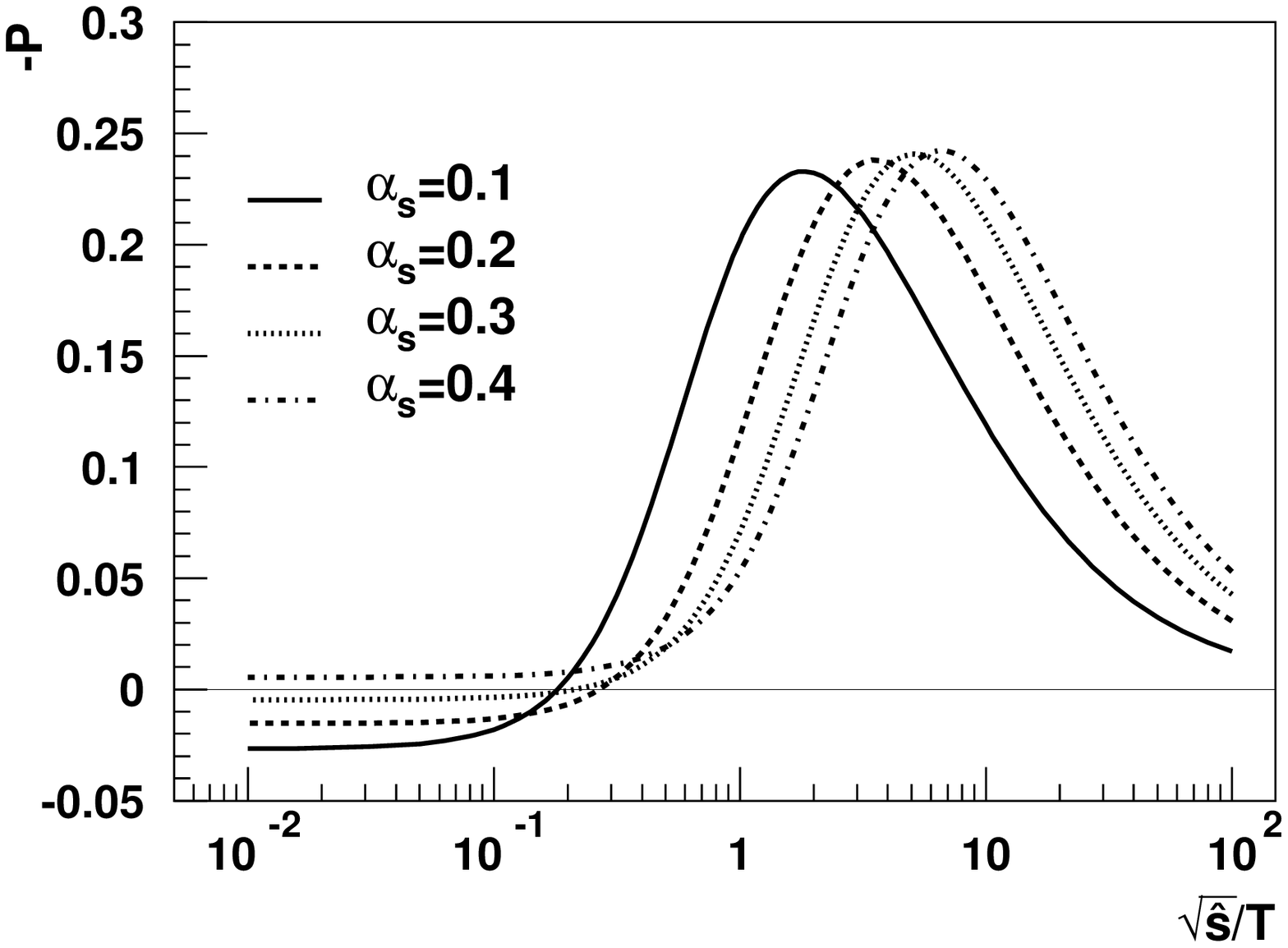}
\includegraphics[width=6.5cm]{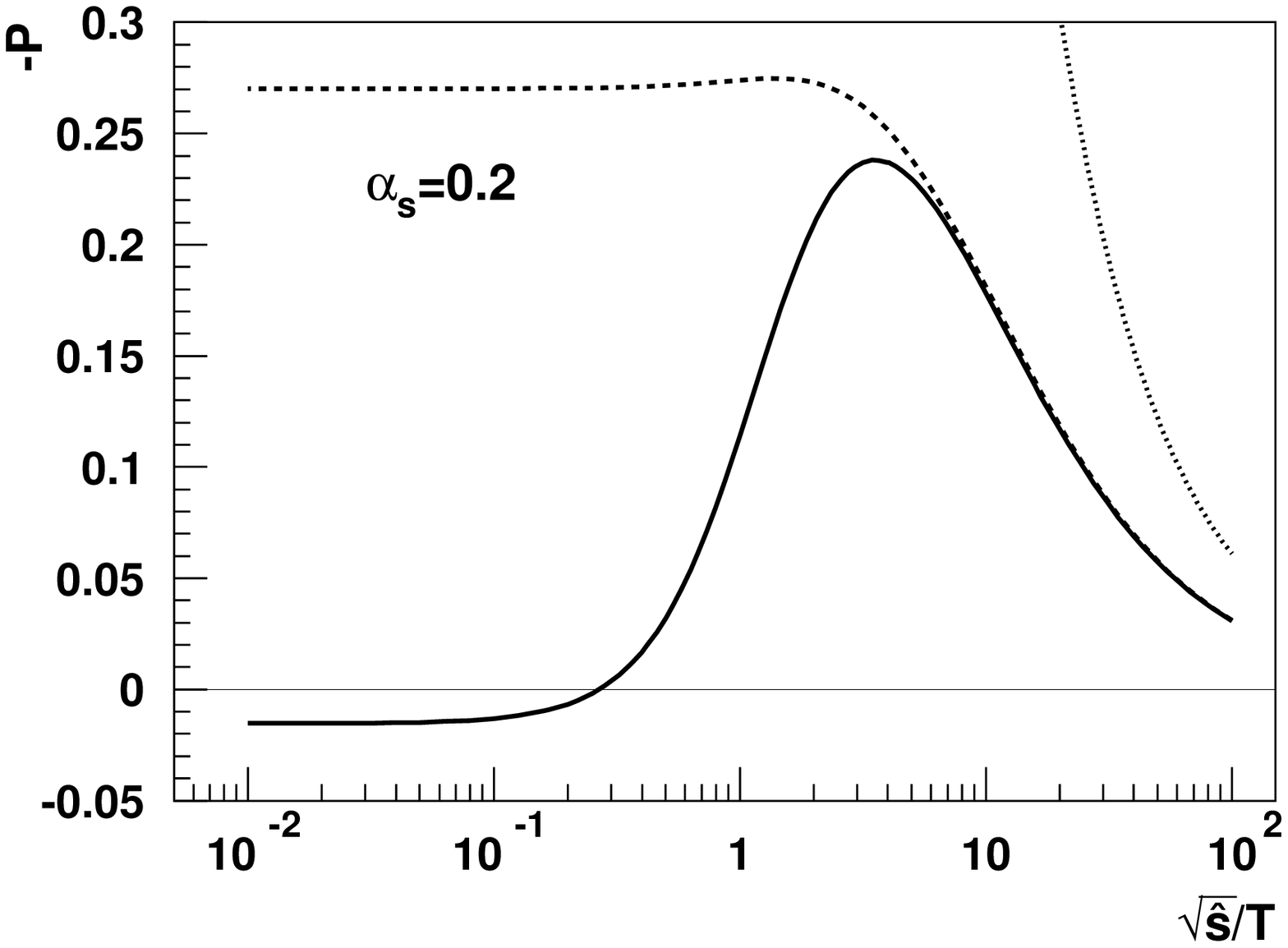}
\caption{a (left), Preliminary results for the 
quark polarization -$P_q$ as a function of
$\sqrt{\hat{s}}/T$ obtained 
using HTL gluon propagator. 
b (right), Comparison with the results obtained under small angle approximation (dashed line) 
and that obtained using static potential model under small angle approximation (dotted line).}
\end{center}
\end{figure}

The results in Fig.~3 show that the polarization is quite different 
for different $\sqrt{\hat{s}}/T$. 
It is very small both in the high energy and low energy limit. 
However, it can be as high as 20\% at moderate $\sqrt{\hat{s}}/T$.  
At RHIC, we can only give a rough estimation of the ratio 
$\sqrt{\hat{s}}/T\sim\Delta p_z/T$ which should be between 0.1 and 2. 
We see that in this range the polarization can be quite significant but can also be only of a few percent. 

From Fig.~3b, we see also that, 
at RHIC energy, small angle approximation 
is not a good approximation\cite{Liang:2004ph}. 
We have to rely on the numerical results obtained 
using HTL gluon propagator without small angle approximation. 
More detailed study including transverse flow of the partons are also underway.

\section{The measurable consequences}

The global polarization of quarks and anti-quarks in QGP should have 
many measurable consequences for the hadrons after hadronization.
The most direct ones are the polarizations of the spin non-zero hadrons. 
Data[8-10] from LEP on $e^+e^-\to Z\to h+X$ tell us that 
polarization of the quark or anti-quark can indeed be transferred to 
final hadrons via hadronization.  
We now present the results for hyperons and vector mesons 
from the global quark polarization discussed above in the following.

\subsection{Hyperon polarization}

For hyperons produced via recombination $qqq\rightarrow H$, 
we obtain, 
\begin{eqnarray}
P_\Lambda=P_s; 
&&\hspace{-5cm}P_\Sigma=(4P_q-P_s-3P_sP_q^2)/(3-4P_qP_s+P_q^2); \\
P_\Xi=(4P_s-P_q-3P_qP_s^2)/(3-4P_qP_s+P_s^2); 
&& 
\end{eqnarray}
We see in particular that $P_H=P_q$ for all the $H=\Lambda$, $\Sigma$ and $\Xi$ if $P_s=P_q$. 

For those produced via the fragmentation $q\to H + X$,
we compare with the longitudinal polarization of hyperons in 
$e^+e^-\to Z^0\to q\bar q\to \Lambda+X$ which has been
measured\cite{lep} and can be explained[11] by assuming
that polarized hyperons contain the initial polarized 
leading quark in its SU(6) wave-function. Similar calculations lead to, 
\begin{eqnarray}
P_\Lambda&=& n_sP_s/(n_s+2f_s), \ \ \ \ 
P_\Sigma=(4f_sP_q-n_sP_s)/3(2f_s+n_s),  \\
P_\Xi &=&(4n_sP_s-f_sP_q)/3(2n_s+f_s), 
\end{eqnarray}
where $n_s$ and $f_s$ are the relative $s$-quark abundances to $u$ and $d$ 
in QGP and fragmentation. 
We see in particular that $P_H=P_q/3$ if $P_s=P_q$ and $f_s=n_s$. 

In dependent of the hadronization mechanisms, 
we expect: 
(1) Hyperons and their anti-particles are similarly polarized;
(2) Different hyperons are also similarly polarized.  
(3) The polarization vanishes in central collisions 
and increases with $b$ in semi-central collisions.
(4) It should have a finite value for small $p_T$ and central 
rapidity but increase with rapidity and eventually 
decreases and vanishes at large rapidities.

\subsection{Vector meson spin alignment}

The polarization of a vector meson $V$ is
described by the spin density matrix $\rho^V$ where 
the diagonal elements $\rho_{11}^V$, $\rho_{00}^V$ and $\rho_{-1-1}^V$
are the relative intensities for the spin component $m$ of $V$ 
to take $1$, $0$, and $-1$ respectively.
$\rho_{00}^V$ can be determined by measuring
the angular distributions of the decay products.
Furthermore, unlike the polarization of hyperons, 
$\rho_{00}^V$ does not know the direction of the reaction plane. 
Therefore, one cannot measure the sign of the quark
polarization through $\rho_{00}^V$ . 
On the other hand, one does not need to determine the
direction of the reaction plane to measure $\rho_{00}^V$.

For vector mesons produced in quark recombination mechanism, we obtain, 
\begin{equation}
\rho^{\rho({\rm rec})}_{00}={(1-P_q^2)}/{(3+P_q^2)}, \hspace{1cm}
\rho^{K^*({\rm rec})}_{00}={(1-P_qP_s)}/{(3+P_qP_s)}.
\end{equation}
We see in particular that $\rho_{00}^V<1/3$ if $V$ is produced in this hadronization scenario. 

For the fragmentation of a polarized quark $q^\uparrow\to V+X$, 
we again compare with $e^+e^-\to Z^0\to q\bar q\to V+X$,
and obtain, 
\begin{equation}
\hspace{-1cm}
\rho^{\rho\ ({\rm frag})}_{00}=\frac{1+\beta P_q^2}{3-\beta P_q^2},
\ \ \ \ 
\rho^{K^*({\rm frag})}_{00}=\frac{f_s}{n_s+f_s}\frac{1
+\beta P_q^2}{3-\beta P_q^2}+
\frac{n_s}{n_s+f_s}\frac{1+\beta P_s^2}{3-\beta P_s^2},
\end{equation}
The parameter $\beta\approx 0.5$ was obtained [12] by 
fitting the $e^+e^-$ data[9,10].
We see that, in this case, $\rho_{00}>1/3$. 
We also see that, in both hadronization scenarios, vector meson spin alignment 
is a $P_q^2$ effect. 
This has the advantage discussed at the beginning of this sub-section 
but also the shortage that it may be very small.

We are happy to know that dedicated efforts have been made in measuring such effects 
at RHIC. See e.g. [13,14].

\section{Summary and outlook}

In summary, we have shown that produced partons have large local
relative orbital angular momentum in non-central $AA$ collisions at high energies. 
Parton scattering with given relative orbital angular momentum 
can polarize quarks along the same direction due to spin-orbital interaction in QCD. 
Such global quark polarization has many measurable consequences 
and the measurements on such effects might open a new window to 
study the properties of QGP in high energy $AA$ collisions.

\section*{Acknowledgments}
I thank the organizers for inviting me to give this talk. 
The material of this presentation is mainly taken from the recent publications [1-3]. 
It is a great pleasure for me to thank the co-authors X.N. Wang, J.H. Gao, 
S.W. Chen and Q. Wang for fruitful collaboration.  
This work was supported in part by NSFC under the No. 10525523.



\end{document}